\documentstyle[preprint,aps]{revtex}
\tightenlines
\begin{document}

\draft

\title{Non equivalent adsorption sites on the ZGB catalytic reaction model}

\author{ \sc {E.R. Fulco and B.C.S. Grandi }} 
\address{Departamento de F\'{\i}sica - Universidade Federal de Santa
Catarina\\ 
88040-900, Florian\'opolis, SC  - Brazil; e-mail:
bartirag@fisica.ufsc.br \\}

\date{\today}

\maketitle

\begin{abstract}

   	We studied in this work a modified version of the Ziff, Gulari and Barshad (ZGB) model in which the adsorption probabilities of $CO$ and $O_2$ are site dependent.  We employed Monte Carlo simulations to determine the phase diagram of the model as a function of the partial pressure of $CO$ in the gaseous phase, and of the relative adsorption probabilities.  For the case where there is no site with affinity only to $O_2$ adsorption, we found that the non physical O-poisoned state disappears.  This feature can not be seen in the usual ZGB model.

\end{abstract}

PACS: 05.70.Ln, 82.65.Jv
\newpage
\vskip2pc    
 
\section{Introduction}
The search for  models to explain the dynamic behavior of nonequilibrium physical systems which present heterogeneous catalytic reactions are the aim of several works in statistical physics [1-9].  The well-known Ziff, Gulari and Barshad model (ZGB model) [10], used to explain the kinetics of reactions like $CO + 1/2 O_2$$\rightarrow$$CO_2$ and $2CO + 2NO$ $\rightarrow$ $N_2 + 2CO_2$ [5,11]on a catalytic surface, shows some features that are not in agreement with the experiments. For example, this model presents, for the catalytic oxidation of $CO$, a continuous phase transition between an active and an O-poisoned states, which is not found experimentally.  Several experiments tell us that the production of $CO_2$ molecules starts as soon as the $CO$ molecules begin to reach the surface [12].  Some attempts were made with the aim to clarify this point [2,5,6,9,12,13].  For instance, Hua and Ma [9] proposed a model where the sites on the catalytic surface are separated in two types: active and not active sites.  In their model, the surface reactions follow the Langmuir-Hinshelwood mechanism, although the adsorption of molecules depends on the kind of the chosen site, i.e., if the site to do the adsorption is active or not active.  They prepared the lattice surface with a fraction $f$ of distributed randomly active sites.  After, they considered that the conditions for the adsorption of molecules on the surface are: (a) an $O_2$ molecule is adsorbed if at least one of the two vacant adjacent sites, randomly chosen, is an active site; (b) a $CO$ molecule is adsorbed with probability $1$ if the empty site, randomly chosen too, is an active site; if the chosen site is not active, the adsorption of $CO$ occurs with probability $p$. They found the phase diagram of their model and showed that the phase transition from the reactive state to the $CO$-poisoned one is a continuous transition for small values of $f$, and it is a discontinuous phase transition when the value of $f$ is large.  When they increase the value of $p$, the reactive region is narrowed.  Besides, they showed that the $O$-poisoned phase does not appear for different values of $f$.   

In this work, we propose a variation of the ZGB model.  We aimed to study the influence in the phase diagram of the hipothesis intoduced by Ziff et al. that all sites at the catalytic surface are equally accessible to adsorption for both incident molecules ($CO$ or $O_2$). In our model, the catalytic surface does not present a site distribution with equal probability for adsorption for both molecules: we proposed that the adsorption of a $CO$ molecule only occurs if the chosen empty site has preference for the adsorption of this molecule, and analogous procedure is followed for the adsorption of $O_2$ molecule.  For this, the lattice presents a fraction $A_{CO}$ of sites that prefers adsorb $CO$ molecules, a fraction $A_O$ of sites that prefers adsorb $O_2$ molecules, and a fraction $A_b$ of sites that accepts both molecules. This model can be understood as a catalytic surface with deffects, where the affinity to adsorption is site dependent.  

	The model was studied by Monte Carlo simulations, and we determined the production rate of $CO_2$ as a function of $A_{CO}$, $A_O$ and of the relative adsorption rate of  CO  molecules, denoted by $y_{\rm CO}$. In this problem, the significant parameter to tell us if the system is in a reactive state is the production rate of $CO_2$, $P_{CO_2}$, because a nonzero number of empty sites does not guarantee that the system is active - it is necessary to know what kind of molecule the vacant sites at the lattice are available to adsorption.  A possible situation to be found is that where we have only one empty site in the lattice, and this site accepts only the adsorption of oxygen: once the $O_2$ molecule needs two vacant sites to be adsorbed, this site always will be empty.  With the purpose to rich our model, we separate the sites in the lattice in three kinds: a fraction $A_{CO}$ of sites that accepts only the adsorption of $CO$, a fraction $A_O$ of sites that accepts only the adsorption of oxygen, and a fraction $A_b$ of sites that accepts the adsorption of both molecules.  The ZGB model would be a particular case of our model, in which the fraction $A_b$ would be $1$. We show that the present model predicts a phase diagram where an absorbing phase  by oxygen atoms does not appear when $A_O = 0.0$ and $A_{CO}$ varies from $0.0$ to $0.8$. In the next Section we present the model, and in Sec. III we present the results obtained for the production rate of $CO_2$, the discussion of the results and conclusions. 

\section{Model and Monte Carlo simulations}

The simplest model to describe the surface catalytic reaction is that one presented by Ziff, Gulari and Barshad [10].  In this model, the catalytic surface is represented by a square lattice, and all sites are accessible to molecules of any kind.  In this surface, molecules of two species are adsorbed: $CO$ and $O_2$ molecules. When a $CO$ molecule is adsorbed, it occupies only one site in the lattice, while the $O_2$ molecule occupies two sites because it dissociates  in two oxygen atoms when it is adsorbed.  If the adsorbed $CO$ has an $O$ atom as its nearest-neighbor, $CO$ and $O$ react forming a $CO_2$ molecule. This new molecule leaves immediately the surface, and two sites are set free in the lattice.  The whole process follows the Langmuir-Hinshelwood mechanism and the following three steps are considered: 

	(1) $\rm CO(g) + V \rightarrow \rm CO(a)$, 

	(2) $\rm O_{2}(g) + 2V \rightarrow 2\rm O(a)$, 

	(3) $\rm CO(a) + \rm O(a) \rightarrow \rm CO_{2}(g) + 2V$,\\
where the labels $a$ and $g$ denote adsorbed and gaseous particles, respectively, and $\rm V$ is a vacant site. The steps (1) and (2) describe the adsorption of the molecules  CO  and  $\rm O_{2}$  , respectively, and the third step represents the proper  reaction between the adsorbed species to form the $\rm CO_{2}$ molecule.  In this model, the vacant sites are free to adsorb both molecules equally. When we take into account that the sites in the lattice are divided into three categories, those steps may be changed:

	(1) $\rm CO(g) + V_{CO,b} \rightarrow \rm CO(a)$, 

	(2) $\rm O_{2}(g) + 2V_{O,b} \rightarrow 2\rm O(a)$, 

	(3) $\rm CO(a) + \rm O(a) \rightarrow \rm CO_{2}(g) + V_{CO,b}V_{O,b}$, \\
where the labels $CO$ and $O$ on $V$ indicates if the vacant site $V$ is a preferential site for a $CO$ or $O_2$ adsorption, respectively.  The label $b$ means no preferential adsorption site.  To describe the whole process we need to consider also the relative adsorption rate of $CO$ molecules, denoted by $y_{CO}$.

	The Monte Carlo simulation  was performed on a square lattice of linear dimension $L$.  For a given value of $L$, the lattice was prepared taking into account the fractions $A_{CO}$, $A_O$ and $A_b$ of sites that accept the adsorption of the molecules. We have used in all of our simulations periodic boundary conditions.  For each value of $y_{CO}$, we choose what kind of molecule will be tried to adsorb.  If the chosen molecule is $CO$, we choose randomly a site $\sigma$ where $CO$ can be adsorbed: if this site is empty and if it accepts to adsorb this molecule, i.e., this is a vacant site of the type $V_{CO}$ or $V_b$, the adsorption is performed; if $\sigma$ is occupied, the trial ends. If the selected molecule is $O_2$, we select randomly two nearest neighbor sites $\sigma_i$ and $\sigma_j$: if both sites are empty and if they are of types $V_O$ or $V_b$, the adsorption happens; if one of them is occupied, or one of them is of $V_{CO}$ type, the trial finishes. Since the adsorption of $CO$ occurs, its four nearest neighbors are analyzed: if one of them is  an $O$ atom, a $CO_2$ molecule is formed, and it leaves immediately the surface.  When this happens, two vacant sites are free for a new adsorption. The same occurs when an $O_2$ molecule is adsorbed: in this case can be formed one or two $CO_2$ molecules, with two or four sites being free in the lattice, respectively. One Monte Carlo step (MCS) equals $L \times L$ adsorption trials. After one MCS, we compute the production number of $CO_2$.  In order to estimate the quantity of interest, we discarded the first $1000$ Monte Carlo steps to reach the stationary state, and we have used the next $6000$ Monte Carlo steps to calculate the averages. 

	In our model, the significant parameter to determine if system is in a reactive or in a non reactive state is the $CO_2$ yielding, $P_{CO_2}$ (in the ZBG model, we can take the fraction of the vacant sites in the lattice).  The parameter $P_{CO_2}$ is used because, even having a nonzero fraction of vacant sites in the lattice, we can not guarantee the state is active: it is necessary to know for what kind of molecules  the vacant sites are available to adsorption.  For example, if in the lattice there is only one vacant site and it does not accept $CO$ molecules for adsorption,  the site will never be occupied.  Figure 1 shows some snapshots of a square lattice with $L=50$, for $y_{CO}=0.45$, $A_{CO}=A_O=0.3$ and $A_b=0.4$, for the inicial configuration (Fig. 1a), after 100 MCS (Fig. 1b) and after 1000 MCS (Fig. 1.c): we can see that the system always presents a number of vacant sites ($V_{CO}$, $V_O$ and $V_b$) different from zero.

\section{Results and Conclusions}

In the ZGB model, the phase diagram shows two absorbing phases (poisoned or inert) and one active (or reactive) phase.  The  transition from the $O$-poisoned phase to the reactive one occurs for $y_{CO}=0.389$, and this is a continuous phase transition.  If the value of $y_{CO}$ increases, the system remains in a reactive phase up to $y_{CO}=0.527$, when  it becomes $CO$-poisoned.  This second transition is a first-order one, where there is an abrupt change in the number of vacant sites for $y_{CO}=0.527$.  However, when the lattice sites present different affinities for the adsorption of molecules, the phase diagram changes.  If $A_{CO}$ or $A_O$ are different from zero, the inert phases ($CO$ or $O$-poisoned phases) disappear.  In the sequence, we present some representative diagrams of the simulations, all the results for a square lattice with $L=50$. 

	Figure 2 shows the results for the production rate of $CO_2$, $P_{CO_2}$, for $A_{CO}=0.0$ and $A_O$  from $0.0$ to $0.8$ (we remember that $A_{CO} + A_O + A_b = 1$).  We can see that $P_{CO_2}$ decreases  drastically when the value of $A_O$ increases, becoming almost zero when $A_O$ is close to $0.8$.  The figure shows that the maximum production of $CO_2$ occurs at different values of $y_{CO}$, and its value decreases when $y_{CO}$ increases; also, when $A_O$ is increased, this maximum  moves in the direction of increasing values of $y_{CO}$.  In the case present, the $CO$-poisoned phase does not occur.  Figure 3 exhibits the production rate of $CO_2$ as a function of $y_{CO}$, for $A_O=0.0$ and $A_{CO}$ assuming values from $0.0$ (ZGB model) to $0.8$.  This is the most interesting case because it is related to disappearance of the non-physical $O$-poisoned phase, which is predicted by the ZGB model.  Also, the character of the transition from the reactive phase to the $CO$-poisoned one changes: the first-order phase transition becomes a continuous transition when the value of $A_{CO}$ is larger than $0.2$. Contrary that observed in Figure 2, the maximum of $P_{CO_2}$ assumes increasing values when $y_{CO}$ also increases.  An intermediate case is that shown in Figure 4: it displays $P_{CO_2}$ as a function of $y_{CO}$, for $A_O = 0.5$ and $A_{CO}$ assuming values in the range $0.1$ to $0.5$.  We can see, in this case, that the system does not present any inert regions, and the maximum production of $CO_2$ is almost constant, only changing the value of $y_{CO}$, for which the maximum occurs. 

	In summary, we have shown that with a slight modification of the ZGB model, which includes a non equal probability of adsorption of the species, it is possible to account for  the experimental result concerning the  absence of $O$-poisoned states  in  the oxidation of $CO$.  We have found the production rate of $CO_2$ as a function of the relative adsorption rate of $CO$ molecules for several conditions of the catalytic surface: the probability of adsorption is site dependent and this fact leads to changes in the phase diagram.  The most interesting case occurs when there is no lattice site with affinity only to $O_2$: the non-physical $O_2$ poisoned state, predicted by the ZGB model, disappears in the present model. We stress that the model can be understood taking into account deffects that modify the chemical properties of the lattice sites.

\section*{Acknowledgements}

B.C.S. Grandi would like W. Figueiredo for useful discussions. This work was supported by the Brazilian agency CNPq.

\begin{figure}

\caption{Snapshots in a square lattice of $L=50$, and for $A_{CO}=A_O=0.3$. (a) Initial configuration. Legend: Up triangle corresponds to the sites that are accessible to the adsorption of $O$, Dark down triangle correponds to the sites that accept only $CO$ molecules, and blank spaces accept both molecules. (b) Lattice after 100 MCS. Legend: Asterisks corresponds to a $CO$ adsorbed molecule, Letter O to a $O$ adsorbed atom and blank space corresponds to a vacant site. (c) Lattice after $1000$ MCS. Legend is the same to (b).}
\end{figure}

\begin{figure}
\caption{Production rate of $CO_2$ as a function of $y_{CO}$, for $A_{CO}=0.0$ and $A_O$ varyng from $0.0$ to $0.8$. The lines serve as a guide to the eyes.}
\end{figure}

\begin{figure}
\caption{Production rate of $CO_2$ as a function of $y_{CO}$, for $A_{O}=0.0$ and $A_{CO}$ varyng from $0.0$ to $0.8$. The lines serve as a guide to the eyes.}
\end{figure}

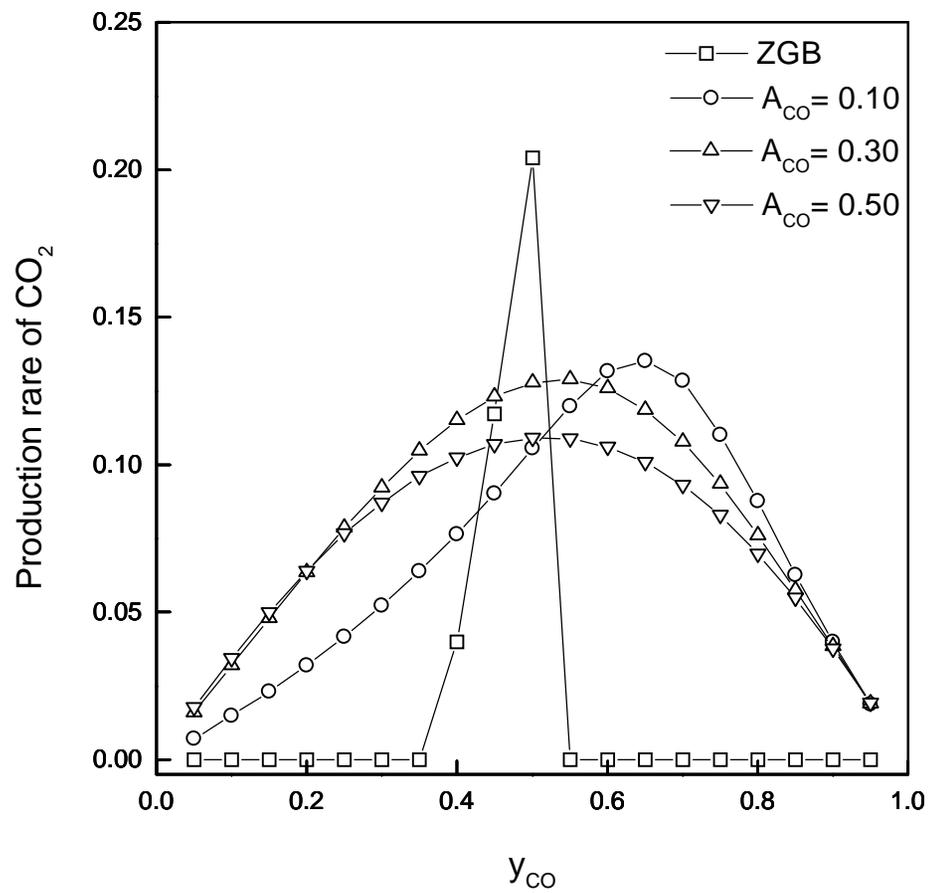
\begin{figure}
\caption{Production rate of $CO_2$ as a function of $y_{CO}$, for $A_O=0.5$ and $A_{CO}$ varyng from $0.0$ to $0.5$. The lines serve as a guide to the eyes. }
\end{figure}


\begin{references}

\bibitem{1} E. V. Albano, Phys. Rev. E {\bf 57}, 6840 (1998).
\bibitem{2} J. Cort\'es, E. Valencia and H. Puschmann,  Phys. Chem. Chem. Phys. {\bf 1}, 1577 (1999).
\bibitem{3} K. M. Khan and N. Ahmed,  Physica A {\bf 280}, 391 (2000).
\bibitem{4} P. I. Hurtado and M. A. Mu\~noz,  Cond. Mat./0004138, april 10, 2000.
\bibitem{5} A. G.  Dickman, B. C. S. Grandi, W. Figueiredo and R. Dickman,  Phys. Rev. E {\bf 29}, 6361 (1999).
\bibitem{6} K. M. Khan, K. Yaldram, J. Khalifeh and M. A. Khan, J. Chem. Phys. {\bf 106}, 8890 (1997).
\bibitem{7} P. L. Krapivsky, Phys. Rev. A {\bf 45}, 1067 (1992).
\bibitem{8} K. Yaldram and M. A. Khan, J. Phys. A: Math. Gen. {\bf 26}, 6135 (1993).
\bibitem{9} Da-yin Hua and Yu-qiang Ma. Phys. Rev. E {\bf 64}, 056102 (2001).
\bibitem{10} R. M. Ziff, E. Gulari and Y. Barshad,  Phys. Rev. Lett. {\bf 56}, 2553 (1986).
\bibitem{11} K. Yaldram and M. A. Khan,  J. Catal. {\bf 131}, 369 (1991).
\bibitem{12} F. Bagnoli, B. Sente, M. Dumont and R. Dagonnier, J. Chem. Phys. {\bf 94}, 777 (1991).
\bibitem{13} V. S. Leite, B. C. S. Grandi and W. Figueiredo, J. Phys A- Math. Gen. {\bf 34}, 1967 (2001).

\end{references}
\end{document}